\documentstyle[12pt]{article}
\newcommand{\Title}[1]{\noi {\Large #1} \\}
\newcommand{\email}[2]{\footnotetext[#1]{e-mail: #2}}%
\newcommand{\foom}[1]{\protect\footnotemark[#1]}
\def\nhq{\hspace{-0.5em}}
\def\nqq{\hspace{-2em}}
\def\al{&\nhq}
\def\lal{&&\nqq}               % left alignment
\def\eql{\al=\al}
\newcommand{\beq}[1]{\begin{equation}\label{#1}}
\newcommand{\eeq}{\end{equation}}
\newcommand{\bear}[1]{\begin{eqnarray}\label{#1}}
\newcommand{\bearr}[1]{\begin{eqnarray}\label{#1}\lal}
\newcommand{\ear}{\end{eqnarray}}
\newcommand{\nn}{\nonumber}
\def\nnv{\nonumber\\[5pt] {}}

\def\nnnv{\nonumber\\[5pt] \lal }
\def\cm{\hspace{1cm}}
\def\inch{\hspace{1in}}
\newcommand{\rf}[1]{(\ref{#1})}
\newcommand{\mcite}[2]{[{#1}-{#2}]}
\newcommand{\nl}{ {\hfill \break} }

\def\noi{\noindent}
\def\yy{\\[5pt]}
\def\yyy{\\[5pt] \lal}

\newcommand{\bdy}{ {\partial } }

\newcommand{\R}{ \mbox{\rm I$\!$R} }

\newcommand{\diag}{ \mbox{\rm diag} }

\newcommand{\sign}{ \mbox{\rm sign} }
\newcommand{\eps}{ \varepsilon }

\newcommand{\M}{ \mbox{\cal M} }

\def\e{{\rm e}}
\def\m{{\rm m}}
\def\half{{\frac{1}{2}}}
\newcommand{\ints}{ \mbox{\rm isc} }
\def\bh{black hole}
\def\bhs{black holes}%
\newcommand\oG{\overline{G}}
\newcommand\oc{\overline{c}}
\newcommand\oI{\overline{I}}
\newcommand\eqdef{\stackrel{\rm def}{=}}

\newcommand{\Acknow}[1]{\subsection*{Acknowledgement} #1}
\def\eq{Eq.\,}
\def\eqs{Eqs.\,}
\def\dst{\displaystyle}
\def\Half{{\dst\frac{1}{2}}}

\def\const{{\rm const}}
\def\DAL{\raisebox{-1.6pt}{\large $\Box$}\,}
\newcommand{\vars}[1]{\left\{\begin{array}{ll}#1\end{array}\right.}
\def\mO{{[-1]\mathstrut}}
\def\pO{{[1]\mathstrut}}
\def\rank{\mathop{\rm rank}\nolimits}
\def\umx{u_{\max}}

\begin{document}
%%%%%%%%%%%%%%%%%%%%%%%%%%%%%%%%%%%%%%%%%%%%%%%%%%%%%%%%%%%%%%%%%%%%%%%
%%% Info page for editor and referees
%%%%%%%%%%%%%%%%%%%%%%%%%%%%%%%%%%%%%%%%%%%%%%%%%%%%%%%%%%%%%%%%%%%%%%%
%\thispagestyle{empty}
%INFO PAGE
%\vspace{1cm}
%\begin{center}
%Title:
%\\
%Authors:
%\\
%\vspace{1cm}
%\end{center}
%Corresponding author:
%\vspace{0.5cm}
%\\
%    Mailing address:
%\\
%                     Dr. Martin Rainer
%\\
%                     Gravitationsprojekt/Kosmologie
%\\
%                     Mathematische Physik I
%\\
%                     Inst. fuer Mathematik
%\\
%                     Universitaet Potsdam
%\\
%                     PF 601553
%\\
%                     D-14415 Potsdam
%\\
%                     Germany
%\vspace{1cm}
%\\
%e-mail: mrainer@aip.de
%\\
%Tel.    + 49 - 331 - 7499280
%\\
%Fax/ES: + 49 - 331 - 2801290
%\\
%\np
%%%%%%%%%%%%%%%%%%%%%%%%%%%%%%%%%%%%%%%%%%%%%%%%%%%%%%%%%%%%%%%%%%%%%%%
%%% Beginning of the paper
%%%%%%%%%%%%%%%%%%%%%%%%%%%%%%%%%%%%%%%%%%%%%%%%%%%%%%%%%%%%%%%%%%%%%%%
\vspace*{-2.0cm}
\centerline{\mbox{\hspace*{10cm} Preprint Math-97/21}}
\centerline{\mbox{\hspace*{10cm} Univ. Potsdam, July 1997%, published in:
}}
\vspace*{1.0cm}
\begin{center}
\Title{\Large\bf Intersecting electric and magnetic $p$-branes:\yy
       spherically symmetric solutions}
\bigskip

\noi{\large\bf K.A. Bronnikov\foom 1\dag, U. Kasper\foom 2\ddag\
        and M. Rainer\foom 3\ddag }

\bigskip
\noi{\dag\ \it Centre for Gravitation and Fundamental Metrology
\\
VNIIMS, 3-1 M. Ulyanovoy St., Moscow 117313, Russia}
\bigskip

\noi{\ddag\
%\it Gravitationsprojekt/Kosmologie, Mathematische Physik I
\it Institut f\"ur Mathematik, Universit\"at Potsdam,
\\
       PF 601553, D-14415 Potsdam, Germany}
\end{center}
\bigskip

%\centerline{Abstract}
{
\noindent
We consider a $D$-dimensional self-gravitating spherically symmetric
configuration of a generalized electro-magnetic $n$-form $F$ 
and a dilatonic scalar field, admitting an interpretation in terms of 
intersecting $p$-branes.
For theories with multiple times, selection rules are obtained,
which obstruct the existence of $p$-branes in certain subspaces.
General static solutions are obtained under a specific restriction on the
model parameters, which corresponds to the known ``intersection
rules". More special families of solutions (with equal charges for 
some of the $F$-field components) 
are found with weakened restrictions on the
input parameters. 
Black-hole solutions are determined, and it is shown that 
in the extreme limit the Hawking temperature may tend to zero, 
a finite value, or infinity, 
depending on the $p$-brane intersection dimension.  
A kind of no-hair theorem is obtained, claiming that black holes
cannot coexist with a quasiscalar component of the $F$-field.
}

%%%%%%%%%%%%%%%%%%%%%%%%%%%%%%%%%%%%%%%%%%%%%%%%%%%%%%%%%%%%%%%%%%%%%%%%
\email 1 {kb@goga.mainet.msk.su}
\email 2 {ukasper@rz.uni-potsdam.de}
\email 3 {mrainer@aip.de}

%%%%%%%%%%%%%%%%%%%%%%%%%%%%%%%%%%%%%%%%%%%%%%%%%%%%%%%%%%%%%%%%%%%%%%%%
\section{Introduction}
\setcounter{equation}{0}
%%%%%%%%%%%%%%%%%%%%%%%%%%%%%%%%%%%%%%%%%%%%%%%%%%%%%%%%%%%%%%%%%%%%%%%%
This paper studies some possible gravitational effects of 
multidimensional unification schemes with hypermembranes,
currently widely discussed as so-called M-theories (see reviews in 
\mcite{1}{5})
%%      \cite{HTW,S,St,D,Gau}
and are closely related to earlier supergravity theories 
\cite{CJS,SS}.
These models contain in their low-energy bosonic sectors sets of
antisymmetric Maxwell-like forms $F$ of various ranks (connected with
highly symmetric, usually flat, subspaces of space-times of 10
and more dimensions), interacting with dilatonic scalar fields.

We discuss static, spherically symmetric systems.
Trying to adhere to the most realistic conditions, we restrict the
consideration to a single $n$-form $F$ (since in 4
dimensions we only deal with a single electromagnetic field),
interacting with a single scalar field, and to ordinary $S^2$ spheres,
although the solution technique is applicable to more general systems.

Nevertheless, we admit the existence of all possible types of
components of $F$-fields compatible with spherical symmetry, 
namely, electric, magnetic and quasiscalar ones. 
It turns out possible to express the general
exact solutions in terms of elementary functions, 
if the input parameters of the model satisfy 
certain orthogonality conditions in minisuperspace. 

These conditions correspond to the known $p$-brane
intersection rules of $M$-theories. 
For the latter many solutions have been obtained 
\mcite{8}{14},
which here coincide with some special cases of our solutions
below.

It is possible to weaken the restrictions upon the
model parameters, and nevertheless to
find special families of solutions,
which have the additional symmetry,
that some $F$-field charges coincide. 
An example is a solution with equal electric and magnetic charges.

Among our solutions, we also select those describing \bhs.
It turns out that a \bh\ cannot coexist with a nonzero quasiscalar
component of the $F$-field. This result generalizes the well-known
no-hair theorems.

The \bh\ solutions depend on 3 integration constants, related to
the electric, the magnetic, and the mass charge. 
It is also shown that the
Hawking temperature of such \bhs\ depends on the intersection dimension
$d_{\ints}$ of the corresponding $p$-branes. In the extreme limit
the \bh\ temperature may tend to zero for $d_{\ints}=0$, a finite limit
for $d_{\ints}=1$, and infinity for $d_{\ints} > 1$.

Similar sets of solutions with a smaller number of integration
constants are obtained for more general models, 
with an additional symmetry, e.g. equal electric and magnetic 
charges.

The paper is organized as follows:
Sect. 2 describes the general model.
Sect. 3 discusses the field equations and defines the 
minisuperspace representation.
Sect. 4 outlines the general construction of solutions
using an orthogonality condition (Sect. 4.1),
and a simplified method (with less restrictions)
for the case of equal charges (Sect. 4.2).
In Sect. 5 singularities and conditions of black holes
are investigated.
Sect. 6 treats more specific electro-magnetic type solutions.
We give a general solution for a special model (Sect. 6.1),
give examples for this solution (Sect. 6.2)
and present a special solution (with equal charges as additional symmetry)
for a more general model (Sect. 6.3).
Sect. 7 concludes with final remarks on the main results.

For convenience we now list some notional conventions 
for indices and their corresponding objects used below:
\nl
\begin{tabular}{lll}
%\item[]
     $L,\ M,\ P \quad$ & $\mapsto $& coordinate labels of the $D$-dimensional 
	Riemannian space $\M$;
\\
%\item[]
     $I,\ J,... \quad$&$ \mapsto $& subsets of $I_0 := \{0,1,\ldots,N\}$;
\\
%\item[]
	$\e,\m \quad $&$  \mapsto $& labels of electric resp. magnetic type 
	forms;
\\
%\item[]
	$s,s' \quad $&$    \mapsto $&  unified indices, $\e I$ or $\m I$;
\\
%\item[]
	$i,\ j,...\quad $&$\mapsto $& labels of subspaces of $\M$;
\\
%\item[]

	$A,\ B,... \quad$&$ \mapsto $& minisuperspace coordinate labels.
\end{tabular}
\nl
As usual, we use the summation convention over repeated indices with
one index in lower the other in upper position.

%%%%%%%%%%%%%%%%%%%%%%%%%%%%%%%%%%%%%%%%%%%%%%%%%%%%%%%%%%%%%%%%%%%%%%%%
\section{The model}
\setcounter{equation}{0}
%%%%%%%%%%%%%%%%%%%%%%%%%%%%%%%%%%%%%%%%%%%%%%%%%%%%%%%%%%%%%%%%%%%%%%%%
We consider a $D$-dimensional classical bosonic field theory with the action
\beq{1}
    S=\int d^Dx \sqrt{g}\left(R - \varphi^{,M}\varphi_{,M} -
         \frac{\eta_F}{n!} F^2 \e^{2\lambda \varphi}\right)
\eeq
where $g=|\det g_{LM}|$, $L,M =0,\ldots ,D-1$, $R$ is the scalar curvature,
$\varphi$ is a scalar matter field, and $\lambda$ is a coupling constant; 
furthermore,
\bear{2}
  F^2 \al \equiv \al F^{M_1,\ldots,M_n}F_{M_1,\ldots,M_n} ,  
\cm
                       n= 2,3,\ldots,D-2 ;                        
\nnv
  F   \eql dU   
\cm \mbox{i.e.} \cm
  F_{M_1,\ldots,M_n}=\bdy_{\,[M_1}U_{M_2,\ldots, M_n]} ,
\ear
where $U$ is a potential $(n-1)$-form and square brackets denote
alternation.  The coefficient $\eta_F=\pm 1$ will be
chosen later to provide a positive energy density of the $F$-field.

The field equations read:
\bearr{3}
   G_M^P\equiv R_M^P-\frac{1}{2}\delta_M^P R=
		T_M^P [\varphi] +  T_M^P [F] ,             
\yyy    \label{4}
   \nabla_M\left(\e^{2\lambda \varphi}
		F^{M M_2,\ldots,M_n}\right)=0,              
\yyy    \label{5}
   \nabla^M\nabla_M\varphi=\eta_F\lambda F^2\e^{2\lambda \varphi},
\ear
where the energy-momentum tensors (EMTs) are
\bear{6}
  T_M^P [\varphi]\eql -\varphi_{,M}\varphi^{,P}
              +\frac{1}{2}\delta_M^P\varphi_{,L}\varphi^{,L}, 
\yy \label{7}
  T_M^P [F]      \eql  \frac{\eta_F}{n!}\e^{2\lambda \varphi}
         \left(-F_{ML_2,\ldots ,L_n}F^{PL_2,\ldots ,L_n}
                      +\half \delta_M^P F^2\right).
\ear

We try to find static, spherically symmetric solutions to the set of
equations \rf{3} to \rf{5}. 

We assume a connected multidimensional
space-time structure with 
\beq{8}
	\M=M_{-1} \times M_0\times M_1\times \cdots \times M_N,
	\cm \dim M_i=d_i, \quad i=0,\ldots, N ,
\eeq
where $M_{-1}\subset {\R}$ corresponds to a radial coordinate $u$,
$M_0 = S^2$ is a 2-sphere, $M_1\subset\R$ is time, and $M_i,\ i>1$
are internal factor spaces.
The metric is assumed correspondingly to be
\bear{9}
     ds^2 \eql
	    \e^{2\alpha(u)}du^2 + \sum_{i=0}^N\e^{2\beta_i(u)}ds_i^2
\nn\\
	    \al\equiv\al 
		-\e^{2\gamma(u)}dt^2+\e^{2\alpha(u)}du^2
              +\e^{2\beta_0(u)}d\Omega^2
              +\sum_{i=2}^N\e^{2\beta_i(u)}ds_i^2  ,  
\ear
where
$ds_0^2 \equiv d\Omega^2=d\theta+\sin^2\theta\, d\phi^2$ 
is the line element on $S^2$,
$ds_1^2 \equiv -dt^2$ with $\beta_1 =: \gamma$,
and $ds_i^2$, $i> 1$, are $u$-independent line elements of internal
Ricci-flat spaces of arbitrary dimensions $d_i$ and signatures
$\eps_i$. 

All fields must be compatible
with spherical symmetry.
Hence we assume $\varphi=\varphi(u)$. 
The $F$-field components may be of electric and
magnetic types. An electric-type component is specified by a
$u$-dependent potential form
\beq{Ue}
	F_{\e I,\,u L_2\ldots L_n} = \bdy_{\,[u}U_{L_2\ldots L_n ]} \cm
	U = U_{L_2,\ldots,L_{n}} dx^{L_2}\wedge\ldots\wedge dx^{L_n}
\eeq
where the coordinate indices $L_j$ belong to a certain subspace
\beq{MI}
	M_I = M_{i_1} \times \cdots \times M_{i_k}
\eeq
of the space-time (\ref{8}), associated with
a subset 
\beq{Ie}
	I = \{i_1,\ldots,i_k\} \subset I_0 \eqdef \{0,1,\ldots,N\}.
\eeq
of the set $I_0$ of possible factor space numbers.
The corresponding dimensions are
\beq{d(I)}
	d(I) \eqdef \sum_{i\in I} d_i,\inch    d(I_0) = D-1.       
\eeq
In the $p$-brane setting \cite{S}, 
one of the coordinates of $M_I$ is time, and the form (\ref{Ue})
describes a $(n-2)$-brane in the remaining subspace of
$M_I$.  By assumption, the subspace $M_0$ does not belong to $M_I$
(that is, $0\not\in I$).

A magnetic-type $F$-form of arbitrary rank $k$ may be defined
as a form dual to some electric-type one, namely,
\beq{Fm}
     F_{\m I,\,M_1\ldots M_k}
	= \e^{-2\lambda\varphi} (*F)_{\e I,\,M_1\ldots M_k}
	\equiv \e^{-2\lambda\varphi}
	\frac{\sqrt{g}}{k!} \eps_{M_1\ldots M_k N_1\ldots N_{D-k}}
	F_{\e I}^{N_1\ldots N_{D-k}},
\eeq
where $*$ is the Hodge operator and
$\eps$ is the totally antisymmetric Levi-Civita symbol. Thus
\beq{11}
	\rank F_{\m I} = D - \rank F_{\e I} = d(\oI)           
\eeq
where $\oI \eqdef I_0 \setminus I$ and nonzero components of $F_{\m I}$
contain indices belonging to the subspace $M_{\oI}$.
Since we are considering a single $n$-form, we must put $k=n$ in
(\ref{Fm}), so that
\beq{12}
	d(I) = n-1   \quad \mbox{for} \quad F_{\e I}, \cm
	d(I) = d(I_0)-n = D-n-1 \quad \mbox{for} \quad F_{\m I}.
\eeq
As before, the subspace $M_0$ does not belong to $M_I$, $0\not\in I$.
So \rf{Fm} describes a magnetic $(D-n-2)$-brane in $M_I$.

Let us label all nontrivial components of $F$ by a collective
index $s = (I_s,\chi_s)$, where $I=I_s\subset I_0$ characterizes the
subspace of $\M$ as described above and
$\chi_s=\pm 1$ according to the rule
\beq{13}
	\e \mapsto \chi_s =+1, \inch \m \mapsto \chi_s = -1.      
\eeq

In both the electric and magnetic cases, the set $I$ either does or
does not include the number 1, refering to the external time
coordinate. If it does, the corresponding $p$-brane evolves with $t$,
and we have a true electric or magnetic field; otherwise
the potential \rf{Ue} does not contain any 4-dimensional
indices and thus behaves as a scalar in 4 dimensions.
In this case we call the corresponding electric-type $F$ component
\rf{Ue} ``electric quasiscalar" and its dual, 
magnetic-type, $F$ component \rf{Fm} ``magnetic quasiscalar". 
So there are in general four types of $F$-field components:
\medskip

A. 	$F_{tuA_3\ldots A_n}$ --- electric ($1\in I$, $A_k$ labeling
	a coordinate of $M_l$, $l\in I$);

B.	$F_{\theta\phi B_3\ldots B_n}$
	                      --- magnetic ($1\in I$, $B_k$ labeling
	a coordinate of $M_l$, $l\in \oI$);

C.	$F_{uA_2\ldots A_n}$ --- electric
	                         quasiscalar ($1\not\in I$, $A_k$ labeling
	a coordinate of $M_l$, $l\in I$);

D.	$F_{t\theta\phi B_4\ldots B_n}$ --- magnetic
	                     quasiscalar ($1\not\in I$, $B_k$ labeling
	a coordinate of $M_l$, $l\in \oI$).
\medskip

The choice of subspaces $I_s$ is arbitrary with the only exception that
any two nontrivial components of $F$ must have at least two different
indices, otherwise there will appear off-diagonal EMT components,
which are forbidden by the Einstein equations, since for our metric the
Ricci tensor is diagonal. Evidently, this is a restriction for
components of the same (electric or magnetic) type, while any electric
component may coexist with any magnetic one. 
Taking this into account, we may
formally consider all
$F_s$ as independent fields (up to index permutations) each with
a single nonzero component.

Let us now pass to the general strategy for solutions, with open
number and types of $F$-field components.
We denote signatures and logarithms of volume factors of the 
subspaces of $\M$ as follows:

\beq{e14}
	\prod_{i\in I} \eps_i =: \eps (I) ;       \cm
	\sum_{i=0}^{N} d_i \beta_i =: \sigma_0 ,  \cm
	\sum_{i=1}^{N} d_i \beta_i =: \sigma_1 ,  \cm
        \sum_{i\in I} d_i \beta_i =: \sigma(I) .   
\eeq
%\end{document}
%%%%%%%%%%%%%%%%%%%%%%%%%%%%%%%%%%%%%%%%%%%%%%%%%%%%%%%%%%%%%%%%%%%%%%%%
\section {Field equations and minisuperspace} % S.3
\setcounter{equation}{0}
%%%%%%%%%%%%%%%%%%%%%%%%%%%%%%%%%%%%%%%%%%%%%%%%%%%%%%%%%%%%%%%%%%%%%%%%
Let us now exploit the possible dimensional reduction
of the present Lagrangian model.
The reparametrization gauge on the lower dimensional manifold
here is chosen as the (generalized) harmonic one.
The variation and the reparametrization gauge 
of spatially homogeneous cosmological 
models can be restricted to the time manifold
(see e.g. \cite{Ra1,Ra2}), for spatially inhomogeneous 
models with homogeneous internal spaces it can be reduced to
a lower dimensional (in the cosmological case space-time) manifold
(see \cite{RZ,Ra}).
In general the dimensional reduction depends on the
symmetry of the problem. Here, due to the general spherical symmetry
and the Ricci-flat internal spaces, the variation reduces to the
radial manifold $M_{-1}$ associated with the radial coordinate, namely $u$.
Then, the harmonic gauge makes $u$ a harmonic coordinate,
as in \cite{Br73}, whence $\DAL u = 0$, such that
\beq{3.1}
	\alpha (u)=\sigma_0 (u).
\eeq
The nonzero Ricci tensor components are then given by
\bear{Ricci}
	\e^{2\alpha}R_t^t   \eql  -\gamma'',                          
\nn\\
	\e^{2\alpha}R_u^u   \eql -\alpha''+{\alpha'}^2-{\gamma'}^2
		       -2{\beta'}^2-\sum_{i=2}^N d_i{\beta'_i}^2,  
\nn\\
     \e^{2\alpha}R_\theta^\theta   \eql
		\e^{2\alpha}R_\phi^\phi= \e^{2\alpha-2\beta}-\beta'',           \nn
\nn\\
     \e^{2\alpha}R_{a_j}^{b_i} \eql
               -\delta_{a_j}^{b_i}\beta''_i \inch   (i,j=1,\ldots,N)\,,
\ear
where a prime denotes $d/du$ and the indices $a_i,\ b_i$ belong to the
$i$-th internal factor space. The Einstein tensor component $G_1^1$ does
not contain second-order derivatives:
\beq{G11}
     \e^{2\alpha}G_1^1=-\e^{2\alpha-2\beta}
    +\Half {\alpha'}^2-\Half \biggl({\gamma'}^2
    +2{\beta'}^2+\sum_{i=2}^N d_i{\beta'_i}^2\biggr).
\eeq
The corresponding component of the Einstein equations is an integral of
other components, similar to the energy integral in cosmology.

The Maxwell-like equations (\ref{4})  are easily solved and give
(with (\ref{3.1})):
\bear{3.2}
	F_{\e I}^{uM_2\ldots M_n}
		\eql Q_{\e I}\e^{-2\alpha - 2\lambda\varphi},
		        \qquad \ \ Q_{\e I}= \const,      
\\\label{3.3}
	F_{\m I,\, uM_1\ldots M_{d(\oI)}}                  
		\eql Q_{\m I} \sqrt{|g_{\oI}|},\qquad
		        \qquad  Q_{\m I}= \const,
\ear
where $|g_{\oI}|$ is the determinant of the $u$-independent
part of the metric of $M_{\oI}$ and $Q_s$ are charges.
These solutions lead to the following form of the EMTs (\ref{7})
written separately for each $F_s$:
\def\Qie{Q_{\e I}}
\def\Qim{Q_{\m I}}
\bearr{3.4}
    \e^{2\alpha}T_M^N [F_{\e I}]
	  = -\half \eta_F \eps(I) \Qie^2 \e^{2y_{\e I}}
		 \diag\bigl(+1,\ \pO_I,\ \mO_{\oI} \bigr);        
\nnnv
    \e^{2\alpha}T_M^N [F_{\m I}]
	  = \half \eta_F \eps(\oI) \Qim^2
	      \e^{2y_{\m I}} \diag\bigl(1,\ \pO_I,\ \mO_{\oI} \bigr),
\ear
where the first place on the diagonal belongs to $u$ and the symbol
$[f]\mathstrut_J$ means that the quantity $f$ takes place on the
diagonal for all indices refering to $M_i,\ i\in J$; the functions
$y_s (u)$ are
\beq{3.5}
	y_s (u) = \sigma (I_s) - \chi_s \lambda\varphi.        
\eeq
The scalar field EMT (\ref{6}) is
\beq{3.6}
	\e^{2\alpha}T_M^N [\varphi]
	    = \half {(\varphi^a)'}^2 \diag\bigl(+1,\ \mO_{I_0}\bigr).
\eeq

The sets $I_s\in I_0$ may be classified by types A, B, C, D according
to the description in the previous section. Denoting $I_s$ for the
respective types by $I_A,\ I_B,\ I_C,\ I_D$, we see from (\ref{3.4})
that, in order to have positive electric and
magnetic energy densities, one has to require
\beq{3.7}
	-\eps(I_A) = \eps(\oI_B) = \eps(I_C) = -\eps(\oI_D) = \eta_F.
\eeq
If $t$ is the only time coordinate, (\ref{3.7}) with $\eta_F=1$
holds for any choices of $I_s$. If there exist other times,
then the relations (\ref{3.7}) are {\em selection rules} for choosing
subspaces where the $F$ components may be specified. 
Especially, they may be of
be of importance in unification theories involving multiple times, see
\cite{Bars}.

Here is an example of how the rules (\ref{3.7}) work. Let there be two
time coordinates $x^0$ and $x^4$ and an electric (A) component of $F$
such that the corresponding subspace $M_{I_A}$ does not include the
coordinate $x^4$ (the electric $p$-brane evolves only with the time
$x^0$). We will express this, by convention, as
$I_A \ni x^0,\ I_A \not\ni x^4$. Then for a magnetic (B) component the
rules (\ref{3.7}) imply that $\oI_B \not\ni x^4$ and consequently
$I_B \ni x^4$. Thus a magnetic $p$-brane must evolve with both times.
In a similar way, for C and D components of the same $F$-field one
easily finds: $I_C \not\ni x^4,\ I_D \ni x^4$.

Returning to the equations, one can notice that each constituent of the
total EMT on the r.h.s. of the Einstein equations (\ref{3}) has
the property
\beq{3.8}
	T_u^u + T_\theta^\theta = 0.                            
\eeq
As a result, the corresponding combination of \eqs(\ref{3})
has a Liouville form and is easily integrated:
\bear{3.9}
	G_u^u + G_\theta^\theta \eql \e^{-2\alpha}
	    \bigl[-\alpha'' + \beta''_0
	       + \e^{2\alpha - 2\beta_0} \bigr] = 0, 
\nnv
	\e^{\beta_0 - \alpha} \eql s(k,u),
\ear
where $k$ is an integration constant (IC) and the function $s(.,.)$
is defined as follows:
\bear{3.10}
	s(k,u) \eqdef \vars{ k^{-1} \sinh kt, \quad & k>0
\\
			                   t,       & k=0
\\
				     k^{-1} \sin kt,     & k<0
                            }
\ear
Another IC is suppressed by adjusting the origin of the $u$
coordinate.

With (\ref{3.9}) the $D$-dimensional line element may be written in the
form
\beq{3.11}
	ds^2 = \frac{\e^{-2\sigma_1}}{s^2(k,u)}
		  \biggl[ \frac{du^2}{s^2(k,u)} + d\Omega^2\biggr]
				+ \sum_{i=1}^{N} \e^{2\beta_i}ds_i^2
\eeq
where $\sigma_1$ has been defined in (\ref{e14}).

We now represent the remaining field equations
in midisuperspace, i.e. in $\sigma$-model form \cite{RZ,Ra}. 
Since our reduced manifold $M_{-1}$ is $1$-dimensional,
here the geometric midisuperspace is in fact just  the minisuperspace
spanned by the $u$-dependent dilatonic scalar fields.
Similar like in \cite{Ra1,Ra2}, we extend this minisuperspace by the 
matter field, thus
treating the whole set of unknowns ${\beta_i(u),\ \varphi(u)}$
as a real-valued vector function $x^A (u)$
in an $(N+1)$-dimensional vector space $V$, so that
$x^A = \beta_A$ for $A=1,\ldots, N$ and $x^{N+1}=\varphi$.
One can then verify that the field equations for
$\beta_i$ and $\varphi$ coincide with the equations of motion
corresponding to the Lagrangian of a Euclidean Toda-like system
\bear{3.13}
	L=\oG_{AB}{x'}^A {x'}^B -  V_Q (y) , \inch
	        V_Q (y) = \sum_s \theta_s Q_s^2 \e^{2y_s} ,
\ear
where $\theta_s$
equals $1$ if $F_s$ is a true electric or magnetic field and otherwise,
if $F_s$ is quasiscalar,  $\theta_s$ equals $-1$, according to (\ref{3.7}).
The nondegenerate, symmetric matrix 
\beq{3.14}
(\oG _{AB})=\pmatrix {
  	G_{ij}&  0      \cr
	0     &  1      \cr }, \inch
			G_{ij} = d_id_j + d_i \delta_{ij}
\eeq
defines a positive-definite metric in $V$.
The energy constraint corresponding to \rf{3.13} is
\beq{3.15}
	E = {\sigma'_1}^2 + \sum_{i=1}^{N} d_i {\beta'}_i^2
	                 +\varphi'^2 + V_Q (y)
	  = \oG_{AB}{x'}^A {x'}^B + V_Q (y)= 2k^2\sign k,
\eeq
whith $k$ from (\ref{3.9}).
The integral (\ref{3.15}) follows here from the ${u \choose u}$
component of (\ref{3}).

The functions $y_s(u)$ (\ref{3.5}) can be represented as scalar
products in $V$ (recall that $s = (I_s, \chi_s)$):
\bear{3.16}
	y_s (u)   = Y_{s,A}  x^A,    \inch
	(Y_{s,A}) = (d_i\delta_{iI_s}, \ \  -\chi_s \lambda),
\ear
where $\delta_{iI} \eqdef \sum_{j\in I}\delta_{ij}$
is an indicator for $i$ belonging to $I$ (1 if $i\in I$ and 0 otherwise).

The contravariant components of $Y_s$ are found using the matrix
$\oG^{AB}$ inverse to $\oG_{AB}$:
\bearr{3.18}
	(\oG ^{AB})=\pmatrix{
		G^{ij}&      0      \cr
		0     &      1      \cr }, \inch
	G^{ij}=\frac{\delta^{ij}}{d_i}-\frac1{D-2} \\ \lal
												 \label{3.19}
	(Y_s{}^A) =
	\biggl(\delta_{iI_s}-\frac{d(I_s)}{D-2}, \ \ -\chi_s \lambda\biggr),
\ear
and the scalar products of different $Y_s$, whose values are of primary
importance for the integrability of our system, are
\beq{3.20}
	Y_{s,A}Y_{s'}{}^A = d(I_s \cap I_{s'})                  
					- \frac{d(I_s)d(I_{s'})}{D-2}
			+ \chi_s\chi_{s'} \lambda^2.
\eeq

%%%%%%%%%%%%%%%%%%%%%%%%%%%%%%%%%%%%%%%%%%%%%%%%%%%%%%%%%%%%%%%%%%%%
\section{Solutions}      % S.4
\setcounter{equation}{0}
%%%%%%%%%%%%%%%%%%%%%%%%%%%%%%%%%%%%%%%%%%%%%%%%%%%%%%%%%%%%%%%%%%%%%%%%
\subsection{Orthogonality}            % SS. 4.1
%%%%%%%%%%%%%%%%%%%%%%%%%%%%%%%%%%%%%%%%%%%%%%%%%%%%%%%%%%%%%%%%%%%%%%%%

The following assumption makes it possible to entirely
integrate the field equations:

\medskip\noi
	{\sl
	The vectors $Y_s$ are mutually orthogonal with
	respect to the metric $\oG_{AB}$, that is,}
\beq{4.1}
	Y_{s,A}Y_{s'}{}^A = \delta_{ss'}N_s^2.
\eeq
(This evidently means that the number of functions $y_s$ does not
exceed the number of equations.) Due to (\ref{12}), the
norms $N_s$ are actually $s$-independent:
\beq{4.1a}
	N_s^2 = d(I_s)\biggl[1- \frac{d(I_s)}{D-2} \biggr] + \lambda^2
	= \frac{(n-1)(D-n-1)}{D-2} + \lambda^2 \eqdef \frac{1}{\nu},
\eeq
$\nu >0$. The orthogonality condition (\ref{4.1}) with (\ref{3.20})
is a special case of a more general integrability condition found
in search for intersecting $p$-brane solutions of Majumdar-Papapetrou
type \cite{IMR}.

Due to (\ref{4.1}), the functions $y_s(u)$ obey the decoupled equations
\beq{4.2}
	y''_s = \theta_s \frac{Q_s^2}{\nu} \e^{2y_s},
\eeq
whence
\beq{4.3}
	\e^{-y_s(u)} = \vars{
	        (|Q_s|/\sqrt{\nu}) s(h_s,\ u+u_s),  & \theta=+1,   \yy
        [|Q_s|/(\sqrt{\nu} h_s)] \cosh[h_s(u+u_s)],\qquad h_s>0, \quad
			                                     & \theta=-1.
			          }
\eeq
where $h_s$ and $u_s$ are ICs and the function $s(.,.)$ was defined in
(\ref{3.10}). For the sought functions $x^A (u)$ we then obtain:
\beq{4.4}
	x^A(u) = \nu \sum_s Y_s{}^A y_s(u) + c^A u + \oc^A,    
\eeq
where the vectors of ICs $c^A$ and $\oc^A$ satisfy the orthogonality
relations $c^A Y_{s,A} = \oc^A Y_{s,A} = 0$, or
\beq{4.5}
	c^i   d_i\delta_{iI_s} - \lambda  c^{N+1}\chi_s =0, \inch
	\oc^i d_i\delta_{iI_s} - \lambda \oc^{N+1}\chi_s=0.       
\eeq
Specifically, the logarithms of the scale factors $\beta_i$ and the
scalar field $\varphi$ are
\bear{4.6}
	\beta_i (u) \eql
	                 \nu \sum_s \biggl[
     \delta_{iI_s} - \frac{d(I_s)}{D-2}\biggr] y_s (u)
			+c^i u + \oc^i,
\\   \label{4.7}
	\varphi (u)\eql
	            - \lambda\nu \sum_s y_s(u) + c^{N+1} u + \oc^{N+1},
\ear
and the function $\sigma_1$ which appears in the metric (\ref{3.11}) is
\bearr{4.8}
	\sigma_1 = -\frac{\nu}{D-2}\sum_s d(I_s)\, y_s(u) + c^0 u + \oc^0
\ear
with
\beq{4.8a}
	c^0 =  \sum_{i=1}^{N} d_i c^i,  \inch
	\oc^0 =  \sum_{i=1}^{N} d_i \oc^i.
\eeq

Finally, the ``conserved energy" $E$ in (\ref{3.15}) is
\beq{4.9}
	E = \nu \sum_s h_s^2\sign h_s + \oG_{AB}c^A c^B        
	                                           = 2k^2 \sign k.
\eeq

The relations (\ref{3.1}), (\ref{3.2}), (\ref{3.3}), (\ref{3.9}),
(\ref{3.11}), (\ref{4.3})--(\ref{4.9}), along with the definitions
(\ref{3.10}) and (\ref{4.1a}) and the restriction (\ref{4.1}), entirely
determine our solution, which is general under the above assumptions.

%%%%%%%%%%%%%%%%%%%%%%%%%%%%%%%%%%%%%%%%%%%%%%%%%%%%%%%%%%%%%%%%%%%%%%%%
\subsection{Coinciding charges}           %% SS. 4.2
%%%%%%%%%%%%%%%%%%%%%%%%%%%%%%%%%%%%%%%%%%%%%%%%%%%%%%%%%%%%%%%%%%%%%%%%

A possible way of integrating the field equations, allowing one to
avoid, at least partially, the orthogonality requirement (\ref{4.1}),
is the assumption that some of the functions $y_s$ coincide. Indeed,
suppose that two functions (\ref{3.5}), say, $y_1$ and $y_2$, coincide
up to a constant addition (which may be then absorbed by re-defining a
charge $Q_1$ or $Q_2$), but the corresponding vectors $Y_1$ and $Y_2$
are neither coinciding, nor orthogonal (otherwise we would have
the previously considered situation). Substituting $y_1\equiv y_2$ into
(\ref{3.16}), one obtains
\beq{4.10}
	(Y_{1,A} - Y_{2,A})x^A =0.                              
\eeq
As all $Y_s$ are constants, this is a constraint reducing the number of
independent unknowns $x^A$. Furthermore, substituting (\ref{4.10}) to
the Lagrange equations for $x^A$, one easily finds:
\beq{4.11}
	-(Y_{1,A} - Y_{2,A}){x''}^A =                          
	\sum_s \theta_s Q_s^2 \e^{2y_s} Y_s^A (Y_{1,A} - Y_{2,A}) =0.
\eeq
In this sum all coefficients of different functions $\e^{2y_s}$
must be zero. Therefore we obtain, first, the orthogonality conditions
\beq{4.12}
	Y_s^A (Y_{1,A} - Y_{2,A}) =0, \cm s \ne 1,2            
\eeq
for the difference $Y_1-Y_2$ and other $Y_s$, and, second, the
following relation for the charges $Q_{1,2}$:
\beq{4.13}
	(\nu^{-1} - Y_1^A Y_{2,A})(\theta_1 Q_1^2 - \theta_2 Q_2^2) =0,
\eeq
where \eq(\ref{4.1a}) is taken into account.
The first multiplier in (\ref{4.13}) is positive ($\oG_{AB}$ is
positive-definite, hence a scalar product of two different vectors with
equal norms is smaller than their norm squared). Therefore
\beq{4.14}
	\theta_1 = \theta_2, \inch Q_1^2 = Q_2^2.             
\eeq

Imposing the constraints (\ref{4.10}), (\ref{4.12}), (\ref{4.14}),
which reduce the numbers of unknowns and integration constants,
one simultaneously reduces the number of restrictions on the input
parameters (by the orthogonality conditions (\ref{4.1})). 
In other
words, a special solution to the field equations may be obtained
with a more general initial model. Due to (\ref{4.14}), this is only
possible when the two components with coinciding charges are of equal
nature:  both must be either true electric/magnetic ones
($\theta_s=1$), or quasiscalar ones ($\theta_s= -1$). The solution
process may continue as described in the previous subsection, so that
the form of the solutions is also similar, but with a reduced number of
variables.  An explicit example is given below.

%%%%%%%%%%%%%%%%%%%%%%%%%%%%%%%%%%%%%%%%%%%%%%%%%%%%%%%%%%%%%%%%%%%%
\section{Singularities and black holes}      % S.5
\setcounter{equation}{0}
%%%%%%%%%%%%%%%%%%%%%%%%%%%%%%%%%%%%%%%%%%%%%%%%%%%%%%%%%%%%%%%%%%%%%%%%
Our solutions generalize the well-known spherically symmetric 
solutions of Einstein
and dilaton gravity (see e.g. \cite{Br95}) and, like these,
combine hyperbolic, trigonometric and power functions, depending on the
signs of the ICs $k$ and $h_s$, so that a considerable diversity of
behaviours is possible. It may be asserted, however, that a generic
solution possesses a naked singularity at the configuration centre,
where $r(u) = \e^{\beta_0} \to 0$. Indeed, without loss of generality,
the range of $u$ is $0 < u < \umx$, where $u=0$ corresponds to flat
spatial infinity, while $\umx$ is finite iff at least one of the
constants $h_s$ is negative, otherwise and $\umx$ is infinite 
(by (\ref{4.9}),
$k<0$ is only possible if some $h_s < 0$). In the former case, $\umx$
is the smallest zero in the set of functions
\beq{5.1}
	\e^{-y_s} \sim \sin [|h_s| (u-u_s)] ,                       
\eeq
whence it is clear from (\ref{4.6}) that, at least
for some of  $i\in\{1,\ldots,n\}$, $\e^{\beta_i} \to\infty$ for $u \to \umx$. 
On the other hand, according to (\ref{3.11}), 
$\sigma_1 \to \infty$, and the coordinate radius shrinks,
\beq{5.2}
	r= \e^{\beta_0} = \e^{-\sigma_1} / s(k,u) \to 0,          
\eeq
provided the denominator is finite. Hence the limit $u\to\umx$
is the centre. Such singularities are similar to the
Reissner-Nordstr\"om repulsive centre, with $g_{tt}\to \infty$
(if $y_s$ in (\ref{5.1}) corresponds to $\theta_s =1$; otherwise
some other $\beta_i$ becomes infinite) 
and diverging energy of the respective $F$-field component.
Possible coincidences of zeros for different $\e^{-y_s}$ do not
essentially alter the situation.

Another generic case is that of $\umx = \infty$, when all $h_s \geq 0$.
Then, as $u\to \infty$, the factors $\e^{\beta_i}$ behave
generically like $\e^{k_i u}$, with constants $k_i$ of either sign,
in general different for different $i$. Therefore again
we have in most cases a naked singularity, but this time it
is not necessarily at the centre. It turns out, however, that
this subclass of solutions can describe \bhs.
So, let us consider the solutions of Subsec. 4.1 and
suppose that all $h_s > 0$ (and hence $k>0$) when all asymptotics are
exponential, and try to select \bh\ (BH) solutions.
(It can be shown that in the case of only some $h_s=0$ there is
no BH solution. A case of interest, when all $h_s=0$, may be
obtained as a limiting one from the subsequent consideration.)

For BHs we require that all $|\beta_i| < \infty$, $i= 2,\ldots,N$
(regularity of extra dimensions), $|\varphi| < \infty$ (regularity of
the scalar field) and $|\beta_0| < \infty$ (finiteness of the spherical
radius) as $u \to \infty$.
With $y_s(u) \sim -h_s u$, this leads to the following constraints on
the ICs:
\beq{5.3}
	c^A = -k \sum_s \Bigl(\delta_{1I_s} + \nu Y_s{}^A h_s\Bigr),										  
\eeq
where $A=1$ corresponds to $i=1$. Then, applying the orthonormality
relations (\ref{4.5})  for $c^A$, we obtain:
\bear{5.4}
	h_s \eql k \delta_{1I_s},
\yy\label{5.5}
	c^A \eql -k \delta^A_1
	+  k \nu \sum_s \delta_{1I_s} Y_s{}^A .             
\ear
Surprisingly, the ``energy condition" (\ref{4.9}) then holds
automatically.

{}From (\ref{5.4}) it is obvious that, if at least one $I_s$ does not include
time ($i=1$), then $h_s =0$, in contrast to our
assumption. Actually $h_s=0$ means that the corresponding $y_s$ has 
power-law asymptotics, uncompensated by exponential asymptotics
of other functions. Therefore we conclude: 
{\em Quasiscalar components of the
$F$-field are incompatible with \bhs.} 
This is a kind of no-hair
theorem for the case of $p$-branes. We have obtained it for the
special case (\ref{4.1}) when the system is integrable, although very
probably it can be proved that the same incompatibility exists
for any values of the input parameters. Such a theorem has been proved
in \cite{Br95} for $D$-dimensional dilaton gravity with any value of
$\lambda$, while the system is integrable only if $\lambda^2 =1/(D-2)$.
On the other hand, one can verify that under the conditions
(\ref{5.4}), (\ref{5.5}) and the additional assumption
$\delta_{1I_s}=1$ (that is, only true electric and magnetic fields are
present), our solutions indeed describe BHs with a horizon at
$u=\infty$. In particular, $g_{tt}\to 0$ as $u\to\infty$ and the light
travel time $t = \int \e^{\alpha-\gamma} du $ diverges as $u\to\infty$.
This family exhausts all BH solutions under the assumptions
made, except maybe the limiting case $k=0$.

In what follows, we restrict ourselves to a field with one true
electric and one true magnetic components and briefly describe the
BH solutions.

%%%%%%%%%%%%%%%%%%%%%%%%%%%%%%%%%%%%%%%%%%%%%%%%%%%%%%%%%%%%%%%%%%%%
\section{Purely electro-magnetic solutions}   % S.6
\setcounter{equation}{0}
%%%%%%%%%%%%%%%%%%%%%%%%%%%%%%%%%%%%%%%%%%%%%%%%%%%%%%%%%%%%%%%%%%%%%%%%
Suppose that there are two $F$-field components, Type A and Type B
according to the classification of Sec. 2. They will be
labelled as $F_\e$ and $F_\m$ and the corresponding sets $I_s\subset
I_0$ as $I_\e$ and $I_\m$. Then a minimal configuration (\ref{8}) of
the space-time $\M$ compatible with an arbitrary choice of $I_s$ has
the following form:

\beq{6.1}
	N=5,\cm I_0 = \{0,1,2,3,4,5\},  \cm
	I_e = \{1,2,3\},        \cm
	I_m = \{1,2,4\},                                        
\eeq
so that
\bearr
	d(I_0) = D-1, \qquad d(I_\e) = n-1, \qquad d(I_\m)= D-n-1,  \qquad
	d(I_\e \cap I_\m) = 1 + d_2;            
\nnnv 
\label{6.2}
	d_1=1, \inch   d_2+d_3 = d_3 + d_5 = n-2.
\ear
The relations (\ref{6.2}) show that, given $D$ and $d_2$, all $d_i$ are
known.

In the ``polybrane'' interpretation [4--7] there is an electric
($n-2$)-brane located on the subspace $M_2\times M_3$ and a
magnetic ($D-n-2$)-brane on the subspace $M_2\times M_4$.
Their intersection dimension $d_{\ints}=d_2$ turns out to be of outmost
importance for the properties of the solutions.

The index $s$ now takes the two values $\e$ and $\m$ and
\bear{6.2a}
	Y_{\e, A} \eql (1, d_2, d_3, 0, 0, -\lambda);\inch
	                 Y_{\m, A} = (1, d_2, 0, d_4, 0, \lambda); 
\nn\\
     Y_\e^A 	\eql (1,1,1,0,0, -\lambda)
	                   - \frac{n-1}{D-2} (1,1,1,1,1,0);       
\nn\\
	Y_\m^A    \eql (1,1,0,1,0, \lambda)
			         - \frac{D-n-1}{D-2} (1,1,1,1,1,0),
\ear
where the last component of each vector refers to
$x^{N+1} = x^6 = \varphi$.

In the solutions presented below the set of
ICs will be reduced by the condition that the
space-time be asymptotically flat at spatial infinity $(u=0)$ and by
a choice of scales in the relevant directions. Namely, we put
\beq{6.3}
	\beta_i(0)=\varphi(0)=0    \cm (i=1,2,3,4,5).           
\eeq
The requirement $\varphi(0)=0$ is convenient and may be always
satisfied by re-defining the charges. The conditions $\beta_i(0)=0$
($i>1$) mean that the real scales of the extra dimensions are hidden in
the internal metrics $ds_i^2$ independent of whether or not they are
assumed to be compact.

%%%%%%%%%%%%%%%%%%%%%%%%%%%%%%%%%%%%%%%%%%%%%%%%%%%%%%%%%%%%%%%%%%%%%%%%
\subsection{General solution for a special model}        % SS. 6.1
%%%%%%%%%%%%%%%%%%%%%%%%%%%%%%%%%%%%%%%%%%%%%%%%%%%%%%%%%%%%%%%%%%%%%%%%
The orthogonality condition (\ref{4.1}) in our case reads:
\beq{6.4}
	\lambda^2=  d_2+1 -\frac{1}{D-2}(n-1)(D-n-1)             
\eeq
Being a relation between the input parameters, this restricts the
choice of the model; but when the model is chosen in this way, the
above solution is general for it.

The solution is entirely determined by the formulae from Subsec. 4.1,
where the quantities (\ref{6.2a}) should be put into
(\ref{4.4}) with $\oc^A=0$ due to (\ref{6.3}):
\beq{6.5}
	x^A(u) = \nu \sum_s Y_s{}^A y_s(u) + c^A u;  \cm   
	\e^{-y_s(u)} = (|Q_s|/\sqrt{\nu}) s(h_s,\ u+u_s).
\eeq
Due to (\ref{6.4}) the parameter $\nu$ is
\beq{6.5a}
	\nu = 1/\sqrt{1+d_2}.                                
\eeq

The constants are connected by the relations
\bearr{6.6}
	\bigl(|Q_{\e,\m}|/\nu\bigr)\, s(h_{\e,\m}, u_{\e,\m}) =1; 
\nnnv
	c^1 + d_2 c^2 + d_3 c^3 - \lambda c^6 =0;  \inch
	c^1 + d_2 c^2 + d_4 c^4 + \lambda c^6 =0;
\nnnv
	\frac{h_{\e}^2\sign h_{\e} + h_{\m}^2 \sign h_{\m}}{1+d_2}
	  + G_{ij} c^i c^j + (c^6)^2  = 2k^2 \sign k,          
\ear
where the matrix $G_{ij}$ is given in (\ref{3.14}) and
all $\oc^A =0$ due to the boundary conditions (\ref{6.3}).
The fields $\varphi$ and $F$ are given by \eqs (\ref{3.2}),
(\ref{3.3}), (\ref{4.7}).

This solution contains 8 independent ICs, namely, $Q_\e,Q_\m,h_\e,h_\m$
and 4 others from the set $c^A$ constrained by \rf{6.6}.
All of them are nontrivial constants, unlike those which may be
absorbed by a rescaling (shifting $\beta_i\to\beta_i+\const$)
or a redefinition of the origin of $u$ ($u\to u+\const$).
It is a direct generalization of the solution
for $D=2n$, $\lambda=0$  obtained in \cite{BrFa} (the so-called
``non-dual" solution for a conformally invariant generalized Maxwell
field), the one for $n=2$ ($D$-dimensional
dilaton gravity) and other previous ones (see \cite{Br95} and
references therein). In particular, in dilaton gravity $n=2,\ d_2=0$
and the integrability condition \rf{6.4} just reads
$\lambda^2 = 1/(D-2)$, which is a well-known relation of string gravity.
This family, however, does not include the familiar
Reissner-Nordstr\"om solution, for which $D=4,\ n=2$, $\lambda=0$,
$d_2=0$ and \eq(\ref{6.4}) does not hold.

In the BH case ((\ref{5.4}), (\ref{5.5}) with $\delta_{1I_s} =1$)
the solution is more transparent after a coordinate transformation
$u\mapsto R$, given by the relation
\beq {6.7}
     \e^{-2ku}=1-2k/R ,
\eeq
which leads to
\bearr{6.8}
     ds^2 = -\frac{1-2k/R}{P_\e^{B}P_\m^{C}} dt^2
		+P_\e^{C}P_m^{B}
     \left(\frac{dR^2}{1-2k/R}+R^2 d\Omega^2\right)
    				+\sum_{i=2}^5 \e^{2\beta_i(u)}ds_i^2 , 
\yyy\label{6.9}
     \e^{2\beta_2} = {P_\e}^{-B}{P_\m}^{-C},
\cm
     \e^{2\beta_3} = \left({P_\m}/{P_\e}\right)^{B},     
\nnnv
     \e^{2\beta_4} = \left({P_\e}/{P_\m}\right)^{C} ,
\cm
     \e^{2\beta_5} = {P_\e}^{C}{P_\m}^{B} ,               
\yyy
\label{6.10}
 e^{2\lambda\varphi} =
     ({P_\e}/{P_\m})^{2\lambda^2/(1+d_2)} ,
 \yyy \label{6.11}
     F_{01M_3\ldots M_n}=-{Q_\e}/{(R^2 P_\e)} ,
\cm
     F_{23M_3 \ldots M_n}=Q_\m \sin\theta,
\ear
with the notations
\bearr{6.12}
	P_{\e,\m} = 1+ p_{\e,\m}/R, \qquad
	p_{\e,\m} = \sqrt{k^2+ (1+d_2) Q_{\e,\m}^2} -k; 
\nnnv
	B = \frac{2(D-n-1)}{(D-2)(1+d_2)}, 
\cm
	C = \frac{2(n-1)}{(D-2)(1+d_2)}.
\ear

The BH gravitational mass as determined from a comparison of \rf{6.8}
with the Schwarzschild metric for $R\to \infty$ is
\beq{6.13}
     G_N M= k + \Half (B p_\e + C p_\m) ,
\eeq
where $G_N$ is the Newtonian gravitational constant.
This expression, due to $k>0$, provides a restriction
upon the charge combination for a given mass, namely,
\beq{6.14}
     B |Q_\e| +  C |Q_\m| < 2G_N M/\sqrt{1+d_2} .
\eeq
The inequality is replaced by equality in the extreme limit $k=0$.
For $k=0$ our BH turns into a naked singularity (at the centre $R=0$)
for any $d_2>0$, while for $d_2=0$ the zero value of $R$ is not a
centre ($g_{22}\neq 0$) but a horizon. In the latter case, if $|Q_e|$
and $|Q_m|$ are different, the remaining extra-dimensional scale
factors are smooth functions for all $R\geq 0$.

The Hawking temperature $T$ of a static,
spherical BH can be found, according to \cite{Wald}, from the relation
\beq{6.15}
     k_{\rm B} T = \kappa/2\pi,           \inch       
     \kappa      =
     (\sqrt{|g_{00}|})'\Big/\sqrt{g_{11}}\biggr|_{\rm horizon}
     =\e^{\gamma-\alpha}|\gamma'|\, \biggr|_{\rm horizon}\,,
\eeq
where a prime, $\alpha$, and $\gamma$ are understood in the sense of the
general metric \rf{8} and $k_{\rm B}$ is the Boltzmann constant.
The expression \rf{6.15} is invariant with respect to radial coordinate
reparametrization, as is necessary for any quantity having a direct
physical meaning. Moreover, it can be shown to be
invariant under conformal mappings if the conformal factor is smooth at
the horizon.

Substituting $g_{00}$ and $g_{11}$ from \rf{6.8}, one obtains:
\beq{6.16}
     T = \frac{1}{2\pi k_{\rm B}} \frac{1}{4k}             
         \left[\frac{4k^2}{(2k+p_\e)(2k+p_\m)}\right]^{1/(d_2+1)}.
\eeq
If $d_2=0$ and both charges are nonzero, this temperature tends to zero
in the extreme limit $k\to 0$;  if $d_2=1$ and both
charges are nonzero, it tends to a finite limit, and in all other cases
it tends to infinity.  Remarkably, it is determined by the
$p$-brane intersection dimension $d_2$ rather than the whole space-time
dimension $D$.

%%%%%%%%%%%%%%%%%%%%%%%%%%%%%%%%%%%%%%%%%%%%%%%%%%%%%%%%%%%%%%%%%%%%%%%%%%%%%%
\subsection{Examples} %SS.6.2
%%%%%%%%%%%%%%%%%%%%%%%%%%%%%%%%%%%%%%%%%%%%%%%%%%%%%%%%%%%%%%%%%%%%%%%%%%%%%%

Let us present some examples of configurations satisfying the
orthogonality condition (\ref{6.4}) with $\lambda=0$. This condition is
then a Diophantus equation for $D$, $n$ and $d_2$. Some of its
solutions are given in the following table, including also the values
of the constants $B$ and $C$ defined in (\ref{6.12}).

$$
\begin{array}{|c|c|c|c|c|c|c|}
\hline                  &&&&&& \\
               & \quad n \quad  & \ d(I_{\e})\ & \ d(I_{\m})\ &
     			   \qquad d_2 \quad & \quad B \quad & \quad C \quad \\
                        &&&&&& \\
      \hline            &&&&&& \\
 D = 4m + 2             &&&&&& \\
\quad  =6,\ 10,\ 14,\   &&&&&& \\
\qquad 18,\ 22,\ 26, \ldots\ & 2m{+}1  & 2m &  2m &\ \ m{-}1 & 1/m  & 1/m\\
                        &&&&&&                             \\
 \qquad D= 11          &  4    &  3 &   6  &    1 & 2/3  &  1/3   \\
		             &  7    &  6 &   3  &    1 & 1/3  &  2/3   \\
			&&&&&&                             \\
\qquad  D=20	        &  7    &  6 &  12  &    3 & 1/3  &  1/6  \\
		             & 13    & 12 &   6  &    3 & 1/6  &  1/3   \\
                        &&&&&&   \\
\hline
\end{array}
$$

Many of these configurations have been discussed in the literature on
M-theory, probably the most well-known one is that of 2- (electric)
and 5- (magnetic) branes intersecting along a string (1-brane) in $D=11$
supergravity.

%%%%%%%%%%%%%%%%%%%%%%%%%%%%%%%%%%%%%%%%%%%%%%%%%%%%%%%%%%%%%%%%%%%%%%%%%%%%%%
\subsection{Special solution for a more general model}  % SS. 6.3
%%%%%%%%%%%%%%%%%%%%%%%%%%%%%%%%%%%%%%%%%%%%%%%%%%%%%%%%%%%%%%%%%%%%%%%%%%%%%%

Let us now cancel the orthogonality condition (\ref{6.4}) (i.e.
consider a more general set of input parameters) but suppose, as in
Subsec. 4.2, $y_\e = y_\m$. As has been shown there, this implies
$Q_\e^2 = Q_\m^2 \eqdef Q^2$.

The charges can be different only in the case $\lambda=d_3=d_4=0$, i.e.
for a conformal field without dilatonic coupling, studied in
\cite{BrFa}, when the electric and magnetic $(n-2)$-branes coincide. In
this and only in this case we have in (\ref{4.10})--(\ref{4.13}) 
$Y_1 =Y_2$.  
Then the charges $Q_\e$ and $Q_\m$ may be arbitrary but enter
into the solution only in the combination $Q_\e^2+Q_\m^2$.

Let us study other cases. We are again work with 
(\ref{6.1})--(\ref{6.2a}). With $y_\e = y_\m \eqdef y(u)$,
\eq(\ref{4.10}) leads to
\beq{6.17}
	d_3\beta_3 -d_4\beta_4 -2\lambda\varphi =0.        
\eeq
Eqs.\,(\ref{4.12}) are irrelevant since we are dealing
with only two functions $y_s$.
The equations of motion for $x^A$ now take the form
\beq{6.18}
	{x^A}'' = Q^2 \e^{2y} (Y^A_{\e} + Y^A_{\m}).       
\eeq
Their proper combination gives $y'' = (1+d_2) Q^2 \e^{2y}$, whence
\beq{6.19}
	\e^{-y} = \sqrt{(1+d_2)Q^2} s (h, u+u_1)          
\eeq
where the function $s(.,.)$ is defined in (\ref{3.10}) and $h,u_1$ are
ICs and, due to (\ref{6.3}),\\ $\sqrt{(1+d_2)Q^2} s (h, u_1)=1$.
Other unknowns are easily determined using (\ref{6.18}) and
(\ref{6.3}):
\bear{6.20}
	x^A \eql \nu Y^A y + c^A; 
\cm
                	Y^A = Y^A_{\e} + Y^A_{\m} = (1, 1, 0, 0, -1, 0);
\\\nn
	\sigma_1 \eql -\nu y + c_0 u.
\ear
Here, as in (\ref{6.5a}), $\nu = 1/(1+d_2)$, but it is now just
a notation. The constants $c_0,\ h,\ c^A\ (A=1,\ldots,6)$ and $k$
(see (\ref{3.9})) are related by

\bear{6.21}
&&    -c^0 + \sum_{i=1}^{5} d_i c^i = 0,       
\qquad      
	c^1 + d_2 c^2 + d_3 c^3 - \lambda c^6 = 0, 
\qquad      
	c^1 + d_2 c^2 + d_4 c^4 + \lambda c^6 = 0,    
\nn\\
&&	2k^2 {\sign k} = {\frac{2h^2 \sign h}{1+ d_2}}
		     (c^0)^2 + \sum_{i=1}^{5} d_i (c^i)^2 + (c^6)^2 .
\ear

This solution contains six independent ICs and, like that of Subsec.
6.1, directly generalizes many previous solutions, including those of
Ref.\cite{BrFa}.  It is valid without restrictions upon the input
parameters of the model. It actually repeats the solutions obtainable
with a single charge, but with a more complicated space-time structure.

The only case when all extra-dimension scale factors
remain finite as $u\to u_{max}$ is again that of a BH. It is
specified by the following values of the ICs:
\beq{6.22}
     k=h>0,\quad c^3=c^4=c^6=0,\quad                          
                             c_2=-c_5=-\frac{k}{1+d_2}\ ,
\quad
     c_0=c^1 =-\frac{d_2 k}{1+d_2}\ .
\eeq
The event horizon occurs at $u=\infty$. After the same transformation
(\ref{6.7}) the metric takes the form
\bear{6.23}
     ds^2_D \eql
            - \frac{1-2k/R}{(1+p/R)^{2\nu}}dt^2
     +(1+p/R)^{2\nu}
                  \left(\frac{dR^2}{1-2k/R}+R^2d\Omega^2\right)   
\nnnv
\inch\cm 
+(1+p/R)^{-2\nu} ds_2^2 +ds_3^2 + ds_4^2 + (1+p/R)^{2\nu}ds_5^2
\ear
with the notation
\beq{6.24}
      p = \sqrt{k^2 + (1+d_2)Q^2}-k.
\eeq
The fields $\varphi$ and $F$ are determined by the relations
\beq{6.25}
     \varphi\equiv 0\ ,
\cm
     F_{01L_3 \ldots L_n} = -\frac{Q} {R^2(1+p/R)}, 
\cm
     F_{23L_3 \ldots L_n} = Q \sin \theta.
\eeq

The mass and the Hawking temperature of such a BH,
calculated as before, are given by the relations
\bear{6.27}
     G_N M = k +p/(1+d_2),  
\inch
     T = \frac{1}{2\pi k_{\rm B}}
	     \frac{1}{4k}\left(\frac{2k}{2k+p}\right)^{2/(d_2+1)}.
\ear
The well-known results for the Reissner-Nordstr\"om metric are
recovered when $d_2=0$. In this case $T\to 0$ in the extreme limit
$k\to 0$.  For $d_2=1$, $T$ tends to a finite limit as $k\to 0$ and
for $d_2>1$ it tends to infinity. As is the case with two
different charges, $T$ does not depend on the space-time dimension $D$,
but depends on the $p$-brane intersection dimension $d_2$.

%%%%%%%%%%%%%%%%%%%%%%%%%%%%%%%%%%%%%%%%%%%%%%%%%%%%%%%%%%%%%%%%%%%%%%%%
\section {Concluding remarks}
\setcounter{equation}{0}
%%%%%%%%%%%%%%%%%%%%%%%%%%%%%%%%%%%%%%%%%%%%%%%%%%%%%%%%%%%%%%%%%%%%%%%%

We have seen that, in a model which may be called the
electro-gravitational sector of M-theory, under certain restrictions
fairly large classes of exact static, spherically symmetric solutions to the 
field equations can be obtained. Trying to be as close as possible to
empirical practice, we restricted ourselves to a treatment of a
single  $F$-form and a 4-dimensional physical space-time.

The main results of possible physical significance are a
non-hair-type theorem for quasiscalar components of an $F$-form and the
behaviour of the BH temperature. The selection rules (\ref{3.7}) for
theories with multiple times are another point of interest.

We have left aside the problem of a physical 4-dimensional
conformal frame, simply treating the 4-metric $g_{\mu\nu}=g_{MN}\quad
(M,N=0,\ldots,3)$ as a physical one. One reason is that the choice of
a physical frame depends on the concrete form of the underlying theory,
whereas this work discusses the weak field limit of a spectrum of
theories, some of them are probably yet to be discovered.
Some more details on this argument may be found in \cite{Br95}. 
Furthermore, the question of the physical frame for effective
(multi-)scalar-tensor theories (e.g. from multidimensional Einstein gravity)
has been discussed in \cite{RZ} (and further Refs. therein), 
concluding that, 
the question of the physical frame is not decidable with certainty
on a purely classical level.  

In any case,
some important features of the solutions are independent
of {\em smooth} conformal transformations of the frame. 
Thus, the BH nature of a solution
and the Hawking temperature are insensitive to conformal factors which
are {smooth} at the horizon.
Furthermore, also the (highly anisotropic) singularities
in non-BH solutions cannot be removed by {smooth} 
conformal transformations.

\Acknow{
The authors 
are grateful for helpful discussions with V. Ivashchuk, V. Melnikov,
and A. Zhuk.
This work was financially supported by
RFBR project grant N 95-02-05785-a,
DFG grants 436 RUS 113/7, 436 RUS 113/236, 
KL 732/4-1, and Schm 911/6. 
It was partially completed 
at Astrophysikalisches Institut Potsdam and
Institut f\"ur Mathematik, Universit\"at Potsdam.
K. B. wishes to express his gratitude to colleagues in Potsdam 
for their kind hospitality.
%Likewise M. R. thanks the Russian Gravitational Society
%for the pleasant atmosphere during his visit there. 
}
%\small


\begin{thebibliography}{99}

\bibitem{HTW}
C. Hull and P. Townsend, "Unity of Superstring Dualities",
{\it Nucl. Phys.\/} {\bf B 438}, 109 (1995),\protect\newline
P. Horava and E. Witten, {\it Nucl. Phys.\/} {\bf B 460}, 506 (1996),
hep-th/9510209; hep-th/9603142.

\bibitem{S}
J.M. Schwarz,  "Lectures on Superstring and M-theory Dualities",
{\it Preprint} ICTP, hep-th/9607201;

\bibitem{St}
K.S. Stelle, "Lectures on Supergravity p-branes ",
hep-th/9701088.

\bibitem{D}
M.J. Duff,  "M-theory (the Theory Formerly Known as Strings)",
{\it Preprint} CTP-TAMU-33/96, hep-th/9608117.

\bibitem{Gau}
J.P. Gauntlett, ``Intersecting Branes", hep-th/970511.
%
%\bibitem{GSW}
%M.B. Green, J.H. Schwarz, and E. Witten, "Superstring
%Theory" in 2 vols. (Cambridge Univ. Press, 1987).

\bibitem{CJS}
E. Cremmer, B. Julia and J. Scherk, {\it Phys. Lett.\/} {\bf B76}, 409
(1978).

\bibitem{SS}
A. Salam and E. Sezgin, eds., "Supergravities in
Diverse Dimensions", reprints in 2 vols., World Scientific (1989).

\bibitem{LPX}
H. L\"u, C.N. Pope and K.W. Xu, "Liouville and Toda Solitons in M-theory",
hep-th/9604058.
 
\bibitem{IMO}
V.D. Ivashchuk and V.N. Melnikov,
{\it Gravitation \& Cosmology} {\bf 2}, No 4 (8), 297 (1996);
hep-th/9612089.

\bibitem{IM2}
V.D. Ivashchuk and V.N. Melnikov,
{\it Phys. Lett.\/} {\bf B 384}, 58 (1996).

\bibitem{IMR}
V. D. Ivashchuk, V. N. Melnikov, and M. Rainer. 
``Multidimensional $\sigma$-models with composite electric 
$p$-branes".
gr-qc/9705005.

\bibitem{AIR}
I.Ya. Aref'eva, M.G. Ivanov and O.A. Rytchkov,
"Properties of Intersecting p-branes in Various Dimensions",
{\it Preprint} SMI-05-97, hep-th/9702077.

\bibitem{AIV}
I.Ya. Aref'eva, M.G. Ivanov and I.V. Volovich, hep-th/9702079.

\bibitem{BrFa}
K.A. Bronnikov and J.C. Fabris,
{\it Gravitation \& Cosmology} {\bf 2}, No 4 (8), 306 (1996)

\bibitem{Ra1}
M. Rainer,
{\it Int. J. Mod. Phys.} {\bf D 4},  397 (1995).

\bibitem{Ra2}
M. Rainer,
{\it Gravitation \& Cosmology} {\bf 1}, 121 (1995).

\bibitem{RZ}
M. Rainer and A. Zhuk, {\it Phys. Rev. }  {\bf D 54} (1996) 6186.

\bibitem{Ra}
M. Rainer, "Effective multi-scalar-tensor theories and $\sigma$-models
{}from multidimensional gravity"
{\it Preprint} Math-97/4, Univ. Potsdam (1997).

\bibitem{Br73}
K.A. Bronnikov, {\it Acta Phys. Polon.} {\bf B 4}, 251 (1973).

\bibitem{Bars}
I. Bars and C. Kounnas, ``Theories with Two Times", hep-th/9703060.

\bibitem{Br95}
K.A. Bronnikov,
{\it Gravitation \& Cosmology} {\bf 1}, 67 (1995).

\bibitem{Wald}
R. Wald, ``General Relativity", Univ. of Chicago Press, Chicago, 1984.

\end{thebibliography}
\end{document}